\documentclass[aps, prl, twocolumn, superscriptaddress, amsmath,  longbibliography]{revtex4-2}
\usepackage[colorlinks=true, citecolor=blue, urlcolor=blue, linkcolor=red]{hyperref}
\renewcommand{\section}[1]{{\par\it #1---}\ignorespaces}
\usepackage{graphicx,orcidlink}

\begin{document}

\title{Axion detection in the non-Markovian noise}
\author{Jing Dong}
\affiliation{School of Physics and Astronomy, Beijing Normal University, and Key Laboratory of Multiscale Spin Physics (Beijing Normal University), Ministry of Education, Beijing 100875, China}
\affiliation{Institute of Physics, Chinese Academy of Sciences, Beijing 100190, China}
\affiliation{University of Chinese Academy of Sciences, Beijing 100049, China}

\author{Shuang-Quan Ma}
\affiliation{School of Physics and Astronomy, Beijing Normal University, and Key Laboratory of Multiscale Spin Physics (Beijing Normal University), Ministry of Education, Beijing 100875, China}

\author{Rong-Hang Chen}
\affiliation{Beijing Computational Science Research Center, Beijing 100193, China}
\affiliation{School of Physics and Astronomy, Beijing Normal University, and Key Laboratory of Multiscale Spin Physics (Beijing Normal University), Ministry of Education, Beijing 100875, China}

\author{Bin-Kai Luo}
\affiliation{School of Physics and Astronomy, Beijing Normal University, and Key Laboratory of Multiscale Spin Physics (Beijing Normal University), Ministry of Education, Beijing 100875, China}

\author{Tao Liu}
\affiliation{School of Physics and Optoelectronics, South China University of Technology, Guangzhou 510640, China}

\author{Qing Ai}
\email{aiqing@bnu.edu.cn}
\affiliation{School of Physics and Astronomy, Beijing Normal University, and Key Laboratory of Multiscale Spin Physics (Beijing Normal University), Ministry of Education, Beijing 100875, China}

\author{Jun-Hong An\orcidlink{0000-0002-3475-0729}}
\email{anjhong@lzu.edu.cn}
\affiliation{Key Laboratory of Quantum Theory and Applications of Ministry of Education, Lanzhou Center for Theoretical Physics, Gansu Provincial Research Center for Basic Disciplines of Quantum Physics, Key Laboratory of Theoretical Physics of Gansu Province, Lanzhou University, Lanzhou 730000, China}

\begin{abstract}
The detection of axion dark matter poses a significant challenge due to its exceptionally weak coupling to standard-model particles. Here we develop a noise-tolerant quantum-metrology scheme based on coupled $^{129}$Xe spin ensembles to search for such weakly interacting particles. Within an experimentally accessible parameter regime, our scheme achieves an ideal exclusion limit on the axion coupling constant that is several orders of magnitude more stringent than that of conventional detection methods. We further investigate the performance of our scheme under dissipative noise and find that its robustness is fundamentally governed by the energy-spectrum structure of the composite spin-environment system. Remarkably, when a bound state emerges in the spectrum, the exclusion limit asymptotically recovers its ideal value in the non-Markovian dynamics, whereas under the Markovian approximation the limit deteriorates significantly due to noise. Our results establish a universal mechanism for noise-resilient exploration of beyond-standard-model physics, which can be practically implemented by engineering bound-state formation via tuning the Larmor frequency of the spin ensembles.
\end{abstract}
\maketitle

\section{Introduction}
Axions were originally proposed to resolve the strong charge-parity problem in quantum chromodynamics (QCD) \cite{peccei1977prl,Peccei1977PRD,Weinberg1978prl,Wilczek1978prl,Kim2010RMP,Kim1979PRL,Abbott1983PLB1,Dine1983PLB,Preskill1983PLB}. Axions and other light pseudoscalar bosons forming axion-like particles stand as compelling candidates for ultralight dark matter \cite{2000PhysRep,Marsh2016PhysRep,yan2025prl,Su2024PRL,Jiang2021NP,Hagmann1990PRD,Brubaker2017PRL,Asztalos2010PRL,DePanfilis1987PRL,Ouellet2019PRL,Caldwell2017PRL,Lawson2019PRL,Alesini2019PRD,Du2018PRL,Eroncel2025PRL,Boutan2018PRL,Ahn2025PRL,Vacalis2025PRL,Kim2024PRL,Lee2020PRL,Darling2020PRL,Berlin2024PRL}. Their mass range spans a broad range from $10^{-21}$ eV to $10^{-3}$ eV \cite{Ernst2018,Schutz2020prd}, which leaves a wide parameter space to explore. Theoretical arguments from lattice QCD \cite{Borsanyi2016Nature} and axion-string networks \cite{Turner1996PR,Youdin1996PRL,Klaer2017JSAP,Arvanitaki2014PRL} further highlight the axion window from 10 $\mu$eV to 1 meV as a preferred regime for experimental searches. A broad suite of experiments has been designed to probe their couplings to photons, gluons, and fermions \cite{Bermudez2010PRL,Moody1984PRD,Sikivie2021RevModPhys,Wu2022PRL,Wu2023PRL,Kumar2026PRD,Ouellet2019PRL,Du2018PRL,Brouwer2022PRD,Millar2017JCAP,Chen2026PRD,Foster2020PRL}. More recently, nuclear magnetic resonance experiments targeting axion-nucleon couplings have attracted growing attention, exploiting the fact that axions can behave as time-oscillating pseudomagnetic fields acting on nuclear spins \cite{Budker2014PRX,Jackson2020,Abel2017PRX,Smorra2019N,Bloch2020,Graham2013PRD,Graham2018PRD,Garcon2019SA,Dong2025PRD,Huang2025PRL,Wang2026Nature,wu2019prl,Aybas2021prl}. Using 720 h of data, a pseudomagnetic field $B_{\rm ac}=(86\pm 140_{\rm stat})$~aT is inferred \cite{Su2024PRL}, whose uncertainty gives the exclusion limit of axions.

Experimental searches for axion dark matter are fundamentally hindered by the extremely weak couplings between axions and standard-model particles, which render the induced pseudomagnetic signals minuscule. In practice, decoherence originating from various noises constitutes another critical bottleneck that suppresses the detection of such faint signals and limits the exclusion limits achievable in axion experiments \cite{Breuer2007}. Previous studies have proposed magnetic amplification schemes for axion-mediated spin-dependent interactions within the axion window, where noise dynamics is approximated under the Markovian framework \cite{wu2019prl,Aybas2021prl,Su2024PRL}. Nevertheless, given the inherently non-Markovian nature of noise-induced decoherence dynamics \cite{deVega2017RMP}, a crucial question is whether the aforementioned performance characteristics are ostensibly caused by the Markovian approximation or fundamentally reflect physical properties. A growing body of research across diverse quantum platforms has demonstrated that non-Markovian effect can serve as a quantum resource to resist destructive impacts of decoherence  \cite{Nemet2016PRA,Liu2011NP,Caldeira1983AP,Weiss1987JLTP,Garg1985JCP,Gilmore2005JPCM,Chin2012PRL,Long2022PRL}. However, the performance of the spin-based detection schemes of axion-like particles under non-Markovian noises remains an open question. Particularly, developing noise-robust detection methods that essentially push axion search sensitivities beyond conventional approaches is also highly desired.

In this work, we propose a noise-robust quantum metrology protocol built upon coupled \(^{129}\)Xe spin ensembles for axion-mediated interactions searches. Operating within experimentally accessible parameter spaces, our protocol can deliver constraints on the axion coupling constant that outperform conventional detection strategies by multiple orders of magnitude under ideal conditions. We examine how dissipative noise affects the measurement performance and reveal that the energy-spectrum properties of the combined spin-environment system determine the intrinsic robustness of our method. A striking effect emerges once a bound state forms inside the energy spectrum: non-Markovian dynamics allow the measurement sensitivity to approach its ideal limit asymptotically. In stark contrast, noise substantially degrades the achievable bounds when the Markovian approximation is adopted. This underlying physical mechanism offers a general pathway toward noise-resistant searches for new physics beyond the standard model, which may be realized experimentally by adjusting the Larmor frequencies of the spin ensembles to create the required bound state.

\begin{figure}[tbp]
\centering
\includegraphics[width=\columnwidth]{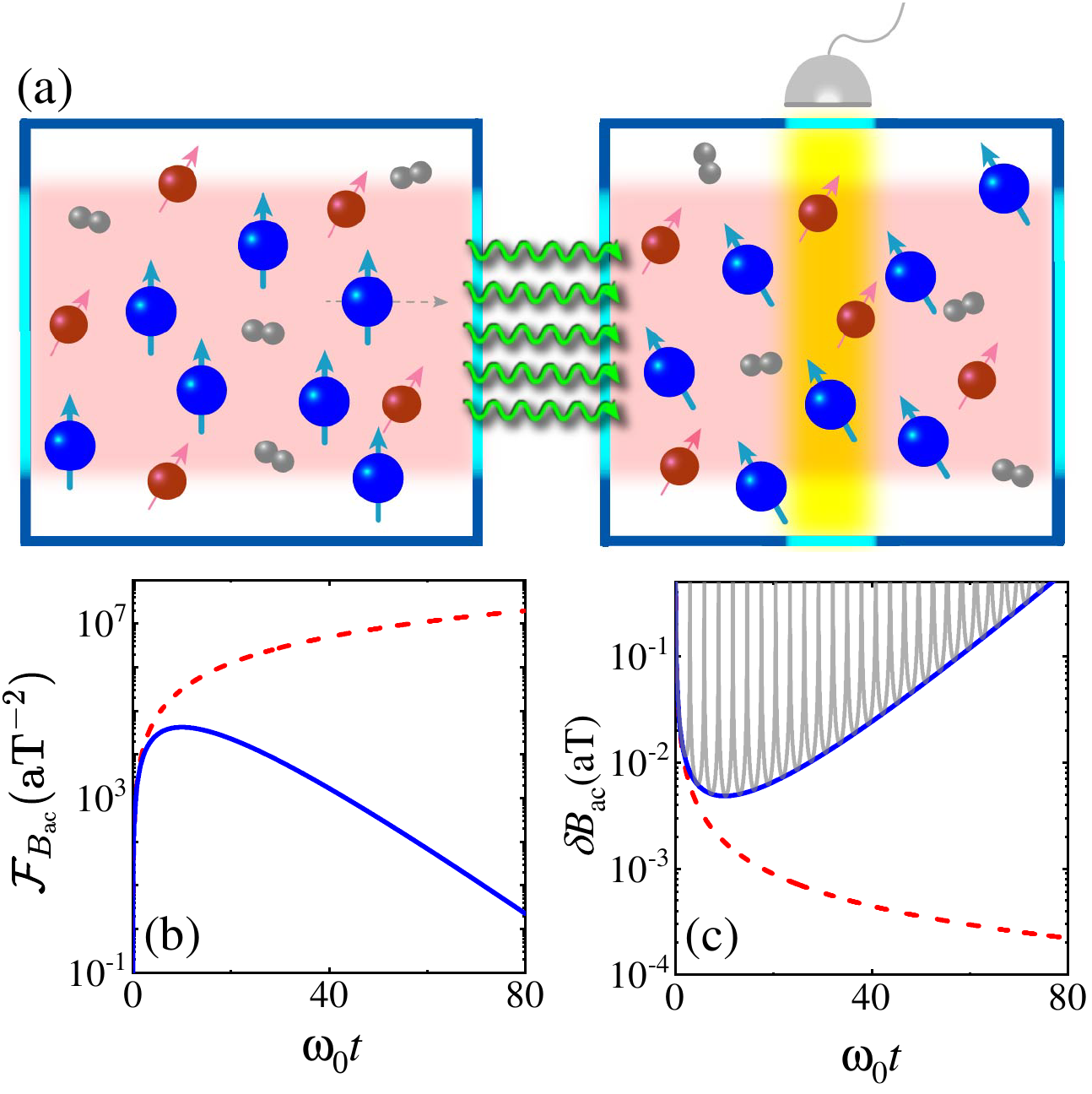}
\caption{(a) Schematic illustration of axion detection. The source cell (left) is tilted into the $x$-$y$ plane by a $\pi/2$ pulse and the nuclear spins therein precess about the $z$ axis, which induces a pseudomagnetic field acting on the $^{129}$Xe spins in the sensor cell (right). Time evolution of (b) the total QFI $\mathcal{F}_{B_\text{ac}}= MN_{\rm se}F_{B_\text{ac}}$ and (c) the pseudomagnetic-field uncertainty $\delta B^F_\text{ac}(t)$ obtained from the QFI in the ideal (red dashed lines) and Markovian-approximate (blue solid lines) cases. $\delta B^P_\text{ac}(t)$ obtained from the error-propagation formula is shown by the gray solid line. We use $M=4.32\times 10^4$,  $N_{\rm se}=2.1\times10^{17}$, $\kappa=0.1\omega_0$, and $B_{\rm ac}=140~{\rm aT}$.}\label{fig:1}
\end{figure}

\section{Ideal axion detection}
To detect the axions, we consider a scheme consisting of two cells, one called the source cell and the other the sensor cell, as shown in Fig.~\ref{fig:1}(a) \cite{Su2024PRL}. Both cells are filled with $^{129}$Xe nuclei, which are initially polarized along the $+z$ axis. The polarization is established through spin-exchange collisions with $^{87}$Rb atoms  \cite{Walker1997,Gentile2017RMP}. The nuclei spins of $^{129}$Xe, which are dominated by the neutron component, are then coupled across the two cells through the mediation of axions. The interaction Hamiltonian reads ($\hbar=c=1$) \cite{Moody1984PRD,Bogdan2006JHEP,Fadeev2019PRA,cong2025RMP}
\begin{align}
V=&-\frac{g_\text{ps}^\text{n}g_\text{ps}^\text{n}}{16\pi m_\text{n}^2}\left[\hat{\pmb{\sigma}}_{\rm so}\cdot\hat{\pmb{\sigma}}_{\rm se}\left(\frac{m_\text{a}}{r^2}+\frac{1}{r^3}\right)-(\hat{\pmb{\sigma}}_{\rm so}\cdot\hat{\mathbf r})\right.\nonumber\\
&\times \left.(\hat{\pmb{\sigma}}_{\rm se}\cdot\hat{\mathbf r})\left(\frac{m_\text{a}^2}{r}+\frac{3m_\text{a}}{r^2}+\frac{3}{r^3}\right)\right]{\rm e}^{-m_\text{a}r},
\end{align}
where $g_\text{ps}^\text{n}g_\text{ps}^\text{n}$ denotes the neutron-neutron pseudoscalar coupling, $m_\text{n}$ and $m_\text{a}$ are the mass of the neutron and the axion, $\hat{\pmb{\sigma}}_{\rm so}$ and $\hat{\pmb{\sigma}}_{\rm se}$ are the neutron spin operators in the source and sensor cells, $\mathbf{r}=r\hat{\bf r}$ is the distance vector between the two interacting neutrons. This axion-mediated interaction $V$ exerts a Zeeman interaction $\gamma_n \mathbf{B}_{\rm p}\cdot\hat{\pmb{\sigma}}_{\rm se}=\gamma_nV/\mu$ on the sensor spins, where $\gamma_n = 2\pi \times 11.78$~Hz/$\mu$T is the gyromagnetic ratio and $\mu$ is the magnetic moment of the $^{129}$Xe nucleus. After a $\pi/2$ pulse, the source spins precess in the $x$-$y$ plane with time. This generates a time-dependent pseudomagnetic field ${\bf B}(t)$. We assume ${\bf B}(t)$ to be along the $y$ axis, i.e., $\mathbf{B}(t)=B_{\text{ac}}\cos(2\pi \nu t)\hat{\bf y}$, where $B_{\text{ac}}=|\mathbf{B}_{\rm p}|N_\text{so}= g_{\rm ps}^{n}g_{\rm ps}^{n}/Q$ and $Q=16\pi m_\text{n}^2\mu {\rm e}^{r m_\text{a}}/N_\text{so}(m_\text{a} r^{-2}+r^{-3})$. $N_\text{so}$ and $\nu$ are the number and the Larmor frequency of source spins of polarized $^{129}$Xe, and $\hat{\bf y}$ is the unit vector of the $y$ axis. The further application of a bias magnetic field ${\bf B}_0=B_0\hat{\bf z}$ to the sensor nuclei results in a Larmor frequency $\omega_0=\gamma_n B_0$ \cite{Jiang2024PNAS}. Therefore, the Hamiltonian of the neutron spin in the sensor cell reads $\tilde{H}_0=\gamma_n[{\bf B}_0+{\bf B}(t)]\cdot{\pmb\sigma}/2$, where the subscript ``se'' has been omitted for brevity. Making the rotating-wave approximation under $\omega_0=2\pi\nu$ and performing a unitary transformation $U=e^{-i\pi\sigma_x/2}e^{ -i \pi (\sigma_x + \sigma_y + \sigma_z)/3\sqrt{3}}$, we obtain $H_0=\omega_A\sigma_z/2$, with $\omega_A=\omega_0-\gamma_nB_\text{ac}/2$. Governed by $H_0$, the initial state $|\psi(0)\rangle=U|+\rangle$ evolves to $|\psi(t)\rangle=e^{i\omega_At/2}[e^{-i\omega_At}|+\rangle+|-\rangle]/\sqrt{2}$, where $\sigma_z|\pm\rangle=\pm|\pm\rangle$.

The axion-mediated interaction is ultimately inferred from the uncertainty of the pseudomagnetic field $\delta B_{\rm ac}$ encoded in the sensor spin \cite{Su2024PRL}. We use quantum Fisher information (QFI) given by $F_{B_{\rm ac}}=|\partial_{B_{\rm ac}} {\bf r}|^2+\left({\bf r}\cdot\partial_{B_{\rm ac}} {\bf r}\right)^2/(1-|{\bf r}|^2)$ \cite{Zhong2013PRA}, where ${\bf r}$ is the Bloch vector of the sensor-spin state $\rho=\frac12(1+{\bf r}\cdot\pmb{\sigma})$, to characterize the maximal information of $B_{\rm ac}$ extractable from $\rho$. According to the quantum Cram\'{e}r-Rao bound, the magnetic-field uncertainty satisfies $\delta B_{\rm ac}^F= 1/\sqrt{M N_{\rm se}F_{B_{\rm ac}}}$ \cite{Helstrom1967PLA,Braunstein1994PRL}, where $M$ is the number of independent measurements, and $N_{\rm se}$ is the number of polarized $^{129}$Xe spins in the sensor cell. A larger QFI corresponds to a smaller field uncertainty and leads to the stronger constraint on the axion-mediated coupling according to the relation between $B_{\rm ac}$ and $g_{\rm ps}^n g_{\rm ps}^n$. It is easy to derive $F_{B_\text{ac}}^{\rm id}(t)={\gamma_n^2 t^2}/{4}$ from $|\psi(t)\rangle$. The magnetic-field uncertainty is $\delta B_{\rm ac}^F(t)=2/(\gamma_nt\sqrt{MN_\text{se}})$.  It provides a baseline for constraining the axion-mediated coupling as \cite{Wu2022PRL,yan2025prl}
\begin{equation}
  g_{\rm ps}^n g_{\rm ps}^n\le Q\delta B_{\rm ac}^F(t)=2Q/(\gamma_nt\sqrt{MN_\text{se}}). \label{idcs}  
\end{equation}
The QFI can be saturated by measuring the operator $\sigma_x$. Then the error propagation formula and the central-limit theorem leads to $\delta B_{\rm ac}^P(t)={\sqrt{1-P(t)^2}\over\sqrt{MN_\text{se}}|\partial_{B_\text{ac}}P(t)|}$, where $P(t)=\langle\psi(t)|\sigma_x|\psi(t)\rangle$. It is easy to find that $\delta B_{\rm ac}^P(t)=\delta B_{\rm ac}^F(t)$, which indicates that measuring $\sigma_x$ is optimal, see Figs. \ref{fig:1}(b) and \ref{fig:1}(c). 

\section{Noisy effects}
However, the inevitable decoherence caused by the environment suppresses the QFI and degrades the performance of axion detection. Determined by whether the system has an energy exchange with the environment or not, the decoherence can be classified into dissipation and dephasing. The dephasing noise \cite{Su2024PRL} can be suppressed by the well-established dynamical decoupling method \cite{PhysRevLett.82.2417,Du2009}. We here investigate the effects of dissipative noise on axion detection. The total Hamiltonian describing the interaction of the sensor spin and a dissipative environment is $H_{\rm tot}= \tilde{H}_{0}+H_{\rm SE}$ with ${H}_{\rm SE}=\sum_k [\omega_k a_k^\dagger a_k+ g_k\sigma_x(a_k+a_k^\dagger)]$, where $a_k$ is the annihilation operator of the $k$th environmental mode with frequency $\omega_k$ and $g_k$ is the coupling strength. In the same procedure of making the rotating-wave approximation and unitary transformation $U$ as the ideal case, we obtain
\begin{equation}\label{eq:Hamiltonian}
H= H_0 + \sum_k [\omega_k a_k^\dagger a_k + g_k(i\sigma_+ a_k+\rm{h.c.})/2].
\end{equation}
The system-environment coupling is characterized by the spectral density $J(\omega)=\sum_kg_k^2\delta(\omega-\omega_k)$. In the continuum limit, it is given as $J(\omega)=\eta\omega(\omega/\omega_c)^{s-1} e^{-\omega/\omega_c}$, where $\eta$ is the coupling strength, $\omega_c$ is the cutoff frequency, and the exponent $s$ classifies the environment into sub-Ohmic for $0 <s< 1$, Ohmic for $ s = 1$, and super-Ohmic for $s > 1$ \cite{Leggett1987RMP}. When the environment is initially in the vacuum state, the total system evolves from $|\Psi(0)\rangle=|\psi(0),\{0_k\}\rangle$ into $|\Psi(t)\rangle = e^{i\omega_At/2}[\alpha(t)\sigma_++\sum_k\beta_k(t)a_k^\dag+1]|-,\{0_k\}\rangle/\sqrt{2}$. Substituting it into the Schr\"{o}dinger equation, we obtain
\begin{equation}\label{eq:s8}
\dot{\alpha}(t) + i\omega_A\alpha(t) +{1\over 4}\int_0^t f(t-\tau)\alpha(\tau)d\tau = 0,
\end{equation}
under $\alpha(0)=1$. $f(t-\tau) = \int_{0}^{\infty}J(\omega)\exp[-i\omega(t-\tau)]d\omega$ is the environmental correlation function. The convolution in Eq.~\eqref{eq:s8} makes the dynamics non-Markovian. A straightforward calculation gives the QFI
\begin{eqnarray}\label{eq:QFI}
F_{B_{\rm ac}}(t)&=&\left|\partial_{B_{\rm ac}}\alpha(t)\right|^2
+\frac{\left[\partial_{B_{\rm ac}}|\alpha(t)|^2\right]^2}{4|\alpha(t)|^2\left[1-|\alpha(t)|^2\right]},
\end{eqnarray}
and the transverse magnetization $P(t)=\text{Re}[\alpha(t)]$. 

When the spin-environment coupling is weak and the environmental correlation time is much shorter than the characteristic time of the spin, we apply the Markovian approximation to Eq.~\eqref{eq:s8} and obtain $\alpha_\text{MA}(t)=\exp\{-[\kappa+i(\omega_A+\Delta)]t\}$, where $\kappa={\pi} J(\omega_A)/4$ is the decay rate and $\Delta={1\over 4}\mathcal{P}\int_0^\infty {J(\omega)\over\omega_A - \omega}d\omega$ is the frequency shift. Here $\mathcal{P}$ denotes the Cauchy principal value. Then the Markovian approximate QFI becomes $F_{B_\text{ac}}^\text{MA}(t)=\gamma _{n}^{2}t^{2}e^{-2\kappa t}/4$, where $\Delta$ has been renormalized into the Larmor frequency $\omega_0$ and the dependence of $\kappa$ on $B_\text{ac}$ has been neglected. It causes the magnetic-field uncertainty, $\delta B_\text{ac}^F(t) = 2e^{\kappa t}/(\gamma_n t \sqrt{M N_\text{se}})$, to grow exponentially and thus the failure of the axion detection. The measured transverse magnetization $P_\text{MA}(t)=e^{-\kappa t}\cos\omega_At$ also loses the information of $B_\text{ac}$ asymptotically, as numerically verified by Figs. \ref{fig:1}(b) and \ref{fig:1}(c). 

In the general non-Markovian case, the asymptotic solution to Eq.~\eqref{eq:s8} can be obtained by the Laplace transform $\tilde{\alpha}(p)=[p+i\omega_A+{1\over 4}\int_0^\infty {J(\omega)\over p + i\omega}d\omega]^{-1}$ of $\alpha(t)$. The inverse Laplace transform to $\tilde{\alpha}(p)$ is performed by finding its poles from
\begin{align}\label{eq:varpi}
y(\varpi) \equiv \omega_A+ \frac{1}{4}\int_{0}^{\infty} \frac{J(\omega)}{\varpi-\omega}d\omega=\varpi,
\end{align}
with $\varpi = ip$. It is interesting to find that the roots $\varpi$ of Eq.~\eqref{eq:varpi} are the eigenenergies of Eq. \eqref{eq:Hamiltonian} in the single-excitation subspace up to a constant $\omega_A/2$. To prove this, we expand the eigenstate as $|\Phi\rangle=(c\sigma_++\sum_k d_ka_k)|-,\{0_k\}\rangle$. Substituting it into $H|\Phi\rangle=E|\Phi\rangle$, we derive $\omega_Ac/2+\sum_k ig_kd_k/2=Ec$ and $-ig_kc/2+(\omega_k-\omega_A/2)d_k=Ed_k$. They readily lead to Eq. \eqref{eq:varpi}, with $E=\varpi-\omega_A/2$, in the continuous limit of the environmental frequencies. Because $y(\varpi)$ is a decreasing function of $\varpi$ in the regime $\varpi < 0$, Eq.~\eqref{eq:varpi} has one isolated root $\varpi_b$ provided $y(0) < 0$. We call the isolated eigenstate with the eigenenergy $\varpi_b$ the bound state \cite{Bai2019PRL}. It has infinite roots in the regime $\varpi >0$ because $y(\varpi)$ is non-analytic. They form a continuous energy band. Applying the inverse Laplace transform by the residue theorem, we have
\begin{equation}\label{eq:ct}
\alpha(t) = Ze^{-i\varpi_b t} + \int_{i\epsilon+0}^{i\epsilon+\infty} \frac{d\varpi}{2\pi} \tilde{\alpha}(-i\varpi)e^{-i\varpi t},
\end{equation}
where $Z = \Big[1 + \frac{1}{4}\int_0^\infty {J(\omega)\over(\omega-\varpi_b)^2} d\omega\Big]^{-1}$. The first term originates from the bound state. The second term arises from the energy band, which tends to zero in the long-time limit due to the out-of-phase interference. Therefore, in the presence of the bound state, we have $\alpha(\infty)= Z\exp(-i\varpi_b t)$, which leads to suppression of decoherence; in the absence of the bound state, $\alpha(\infty)=0$, indicating complete decoherence. It is evaluated from $y(0)<0$ that the bound state is formed for the Ohmic-family spectral density when $\omega_A<\eta\Gamma(s)\omega_c/4$, where $\Gamma(s)$ is the Euler’s gamma function. The result reveals that one can qualitatively manipulate the spin dynamics by tuning its Larmor frequency $\omega_0$ to form the bound state. It is natural to expect that the axion detection fails in the same manner as the the Markovian approximate case when the bound state is absent. We focus on the case with the bound state. We calculate the long-time QFI and the magnetization as
\begin{eqnarray}
\label{eq:FBac} \lim_{t\rightarrow\infty}F_{B_{\rm ac}}(t)&=&\frac{\gamma_n^2t^2Z^4}{4}+{2-Z^2\over 1-Z^2}(\partial_{B_\text{ac}}Z)^2,\label{eq:FBS_longtime}\\ 
\lim_{t\rightarrow\infty}P(t)&=&Z\cos \varpi_bt,
\end{eqnarray}
where $\partial_{B_\text{ac}}\varpi_b=-\gamma_nZ/2$ derived from Eq. \eqref{eq:varpi} has been used. It is remarkable to find that the bound state restores the long-time accumulation of field information that is lost under the Markovian approximation. This is the central mechanism by which the non-Markovian effect improves the sensitivity of the weak axion-induced pseudomagnetic field under realistic noisy conditions.

\begin{figure}[tbp]
\centering
\includegraphics[width=8.5cm]{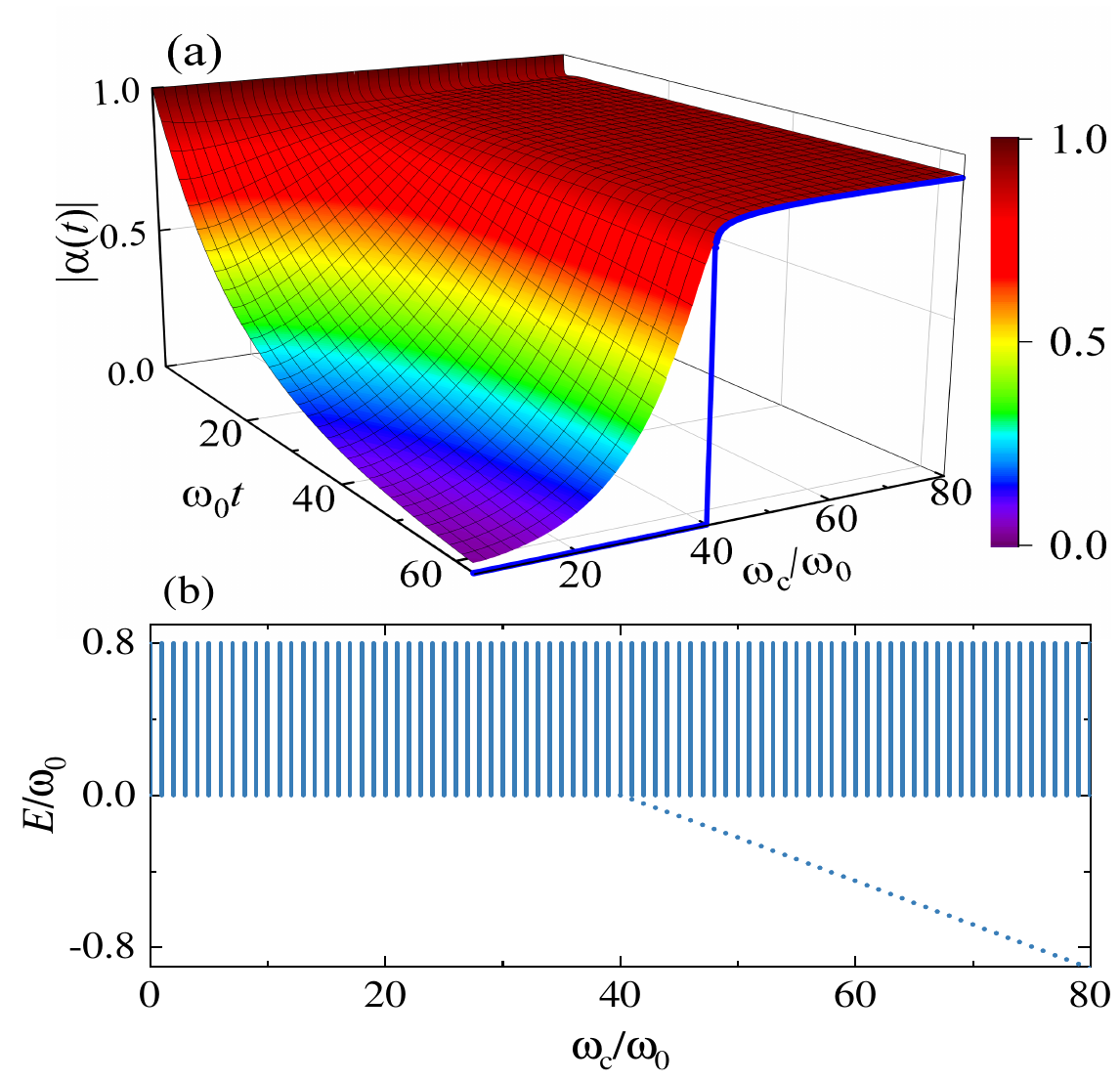}
\caption{(a) Time evolution of $|\alpha(t)|$ as a function of the cutoff frequency $\omega_c$. The blue solid line represents $Z$ calculated from the bound-state eigenenergy $\varpi_b$, which coincides with $|\alpha(\infty)|=Z$. (b) Energy spectrum of the whole system, featuring a continuum band and a branch of discrete bound states isolated within the band gap. The parameters are $s=1$, $\eta=0.1$, and $B_{\rm ac}=140$~aT.}\label{fig:2}
\end{figure}

\section{Numerical simulation}
\begin{figure}[tbp]
\centering
\includegraphics[width=\columnwidth]{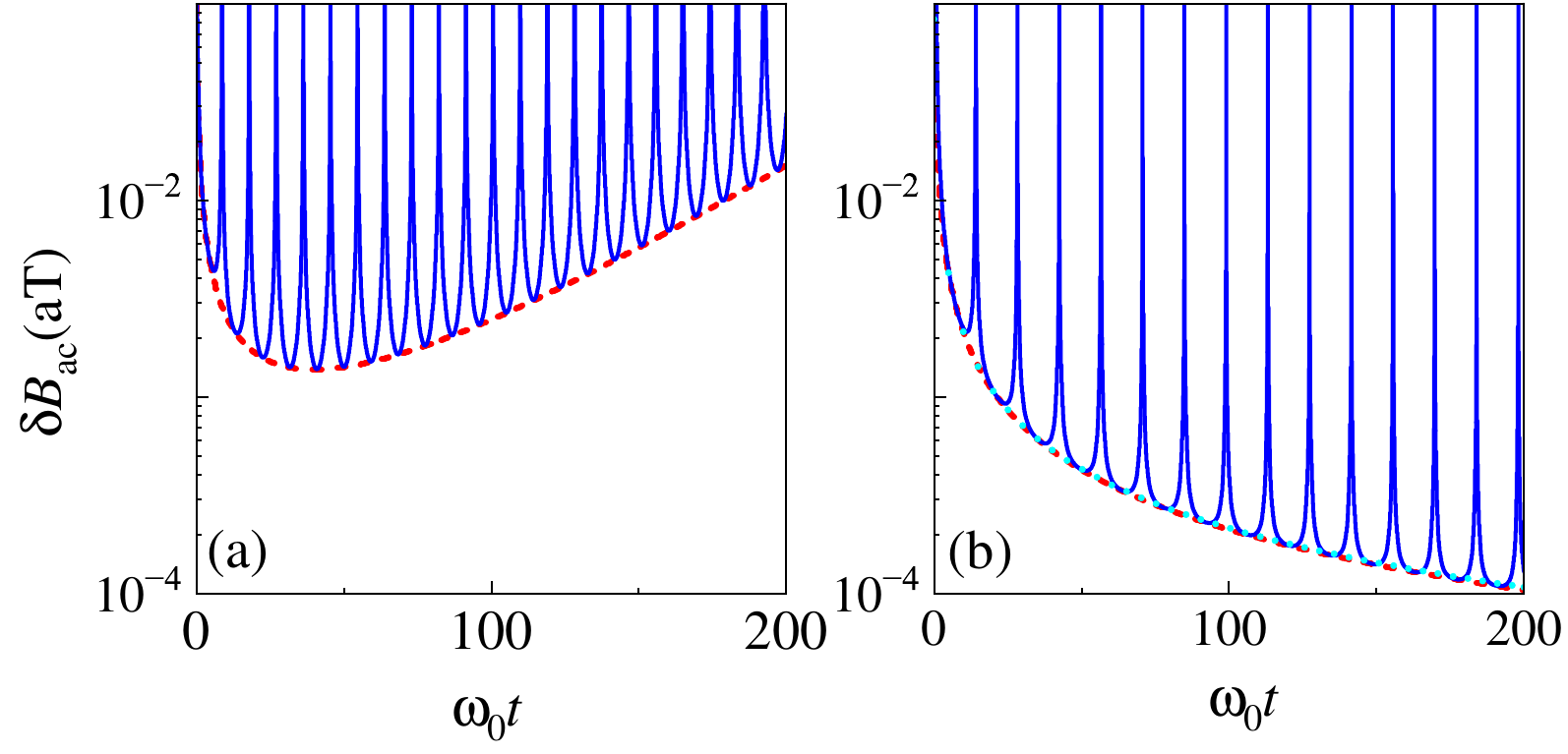}
\caption{Pseudomagnetic-field uncertainty $\delta B_{\rm ac}^P(t)$ (blue solid lines) obtained from measuring transverse polarization $P(t)$ when (a) $\omega_c=25\omega_0$ and (b) $50\omega_0$. The red dashed lines represent the quantum Cram\'{e}r-Rao bound determined by the QFI, which matches with the solution of Eq.~\eqref{eq:FBac} (cyan line). Besides $M=4.32\times 10^4$ and $N_{\rm se}=2.1\times10^{17}$, we use the same parameters values as Fig. \ref{fig:2}.}\label{fig:3}
\end{figure}

To verify the role of the bound state in the decoherence dynamics, we numerically solve Eq.~\eqref{eq:s8} for an Ohmic spectral density and obtain the evolution of $|\alpha(t)|$. For $s=1$, the bound state is present when $\omega_c>4\omega_A/\eta$. We find from Fig.~\ref{fig:2}(a) that $|\alpha(t)$ tends to finite values matching $Z$ when this condition is satisfied. Figure~\ref{fig:2}(b) shows the energy spectrum of the total system by solving Eq. \eqref{eq:varpi}. The result that the regime in the presence of the bound state corresponds exactly to the one where $\alpha(t)$ tends to finite values clearly verifies the dominant role played by the bound state in the decoherence dynamics of the spin. Figure~\ref{fig:3} shows the time evolution of the pseudomagnetic-field uncertainty $\delta B_{\rm ac}^F(t)$ obtained from the QFI via Eq.~\eqref{eq:QFI} and $\delta B_{\rm ac}^P(t)$ from measuring the transverse polarization $P(t)$. The envelope formed by the local minima of $\delta B_{\rm ac}^P(t)$ coincides with $\delta B_{\rm ac}^F(t)$ over the relevant evolution time. It demonstrates that the measurement of the transverse polarization still saturates the QFI and therefore attains the optimal precision allowed by quantum parameter-estimation theory. A comparison between Figs. \ref{fig:3}(a) and \ref{fig:3}(b) highlights the crucial role of the bound state in determining the performance of pseudomagnetic-field detection. In the absence of a bound state, as shown in Fig. \ref{fig:3}(a), the magnetic-field uncertainty $\delta B_{\rm ac}(t)$ initially decreases and reaches a minimal at a finite measured time, but subsequently increases and eventually diverges. Being qualitatively consistent with the Markovian approximate result, this loss of sensitivity results in a relatively small exclusion region of the coupling $g_\text{ps}^\text{n}g_\text{ps}^\text{n}$. However, it is interesting to see that $\delta B_{\rm ac}(t)$ monotonically decreases with time when the bound state exists, matching to the value evaluated from Eq. \eqref{eq:FBS_longtime}, as shown by the cyan dotted line in Fig. \ref{fig:3}(b). Therefore, $\delta B_{\rm ac}(t)$ can asymptotically approach zero as the interaction time increases, correspondingly pushing the attainable exclusion limit on the coupling $g_\text{ps}^\text{n}g_\text{ps}^\text{n}$ toward arbitrarily weak values. 

\begin{figure}[tbp]
\centering
\includegraphics[width=0.9\columnwidth]{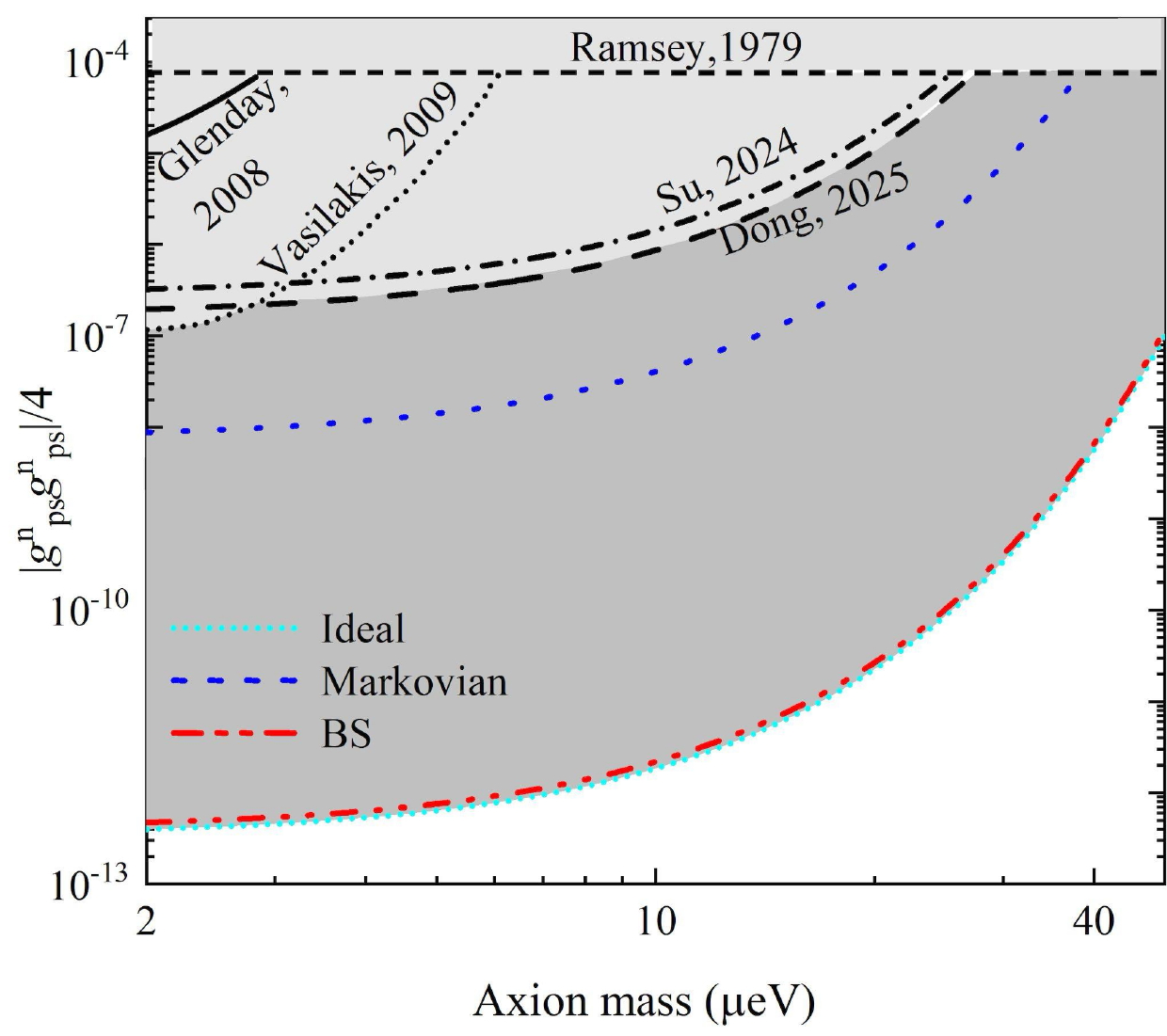}
\caption{\label{fig:4}Constraints on the coupling strength $|g_{\rm ps}^{n}g_{\rm ps}^{n}|/4$ as a function of axion mass. The cyan dotted, blue dotted curves, and red double-dot-dashed denote the ideal, Markovian, and bound-state cases, respectively. The black curves show constraints from the experimental results of Ramsey \cite{Ramsey1979PhysicaA}, Glenday \textit{et al.} \cite{Glenday2008PRL}, Vasilakis \textit{et al.} \cite{Vasilakis2009PRL}, and Su \textit{et al.} \cite{Su2024PRL}, and  theoretical result of Dong \textit{et al.} \cite{Dong2025PRD}.
We use the same parameter values as Fig. \ref{fig:2} except for $\omega_c=50\omega_0$ and $t=100\omega_0^{-1}$.}
\end{figure}

Figure~\ref{fig:4} shows the constraints on $|g^n_{\rm ps}g^n_{\rm ps}|/4$ when $t=100\omega_0^{-1}$, together with the existing limits of previous works \cite{Ramsey1979PhysicaA,Glenday2008PRL,Vasilakis2009PRL,Su2024PRL,Dong2025PRD}. The ideal noise-free evolution evaluated from Eq. \eqref{idcs} gives the most stringent exclusion limit, which is much smaller than the previous results. In the presence of dissipative noise, the decoherence in the Markovian-approximation case suppresses the available QFI and therefore leads to an increased exclusion limit on $|g_{\rm ps}^n g_{\rm ps}^n|/4$. This limit becomes larger with increasing $t$, which further deteriorates the detection performance. It is remarkable to find that,  according to Eq. \eqref{eq:FBS_longtime}, the exclusion limit asymptotically approaches
\begin{equation}
|g_\text{ps}^ng_\text{ps}^n|\le 2Q/(Z^2\gamma_nt\sqrt{MN_\text{se}})
\end{equation}
in the non-Markovian case when the bound state is present. It recovers the ideal result \eqref{idcs} up to a factor $1/Z^2$, see the dot-dot-dashed line of Fig. \ref{fig:4}. It is interesting to find that our scheme sets the most stringent record of the exclusion limit of $|g_{\rm ps}^{n}g_{\rm ps}^{n}|/4$, being $10^6$ times smaller than previous measurements near $m_a=3.5~\mu$eV  \cite{Su2024PRL}. These results demonstrate that the non‑Markovian effect and the spin-environment bound state supply a fundamental mechanism to noise-tolerant high-precision detection of axion dark matter.

\section{Discussions and conclusion}
Although our numerical calculation is performed by changing $\omega_c$, a same conclusion can be drawn by changing the Larmor frequency $\omega_0$ because only the dimensionless ratio $\omega_0/\omega_c$ governs the formation of the bound state. Thus, when the cutoff frequency $\omega_c$ is fixed, there remains a tunable window of $\omega_0$ that supports the bound state. Operating within this window can equivalently preserve the ideal performance of the axion detection. While we adopt the Ohmic spectral density, our bound-state mechanism is applicable in other forms, where the explicit bound-state condition differs, but the underlying mechanism remains unchanged. Therefore, once the actual spectral density of the noise is characterized, our scheme can be immediately applied. The bound state and its dynamical effect have been experimentally validated in circuit QED systems \cite{Liu2017} and ultracold-atom systems \cite{nature5,Kwon2022,Kim2025}. These progresses give a strong support to the physical principle of our noise-tolerant axion detection.

In conclusion, we have developed a noise-tolerant scheme for detecting axion-mediated interactions with coupled $^{129}$Xe spin ensembles in the quantum-metrology framework, whose ideal exclusion limit of the axion is smaller than that of conventional schemes by several orders of magnitude in an experimentally accessible parameter regime. We have found that its performance in the presence of dissipative noise is essentially determined by the feature of energy spectrum formed by the spin and its environment. When a bound state is present in the energy spectrum, the exclusion limit asymptotically restores its ideal value in the non-Markovian dynamics. This is in sharp contrast to the Markovian approximate case, where the exclusion limit become deteriorated by the noise. Our result gives a universal mechanism to develop noise-tolerant exploration of weak signals beyond the Standard Model by engineering the formation of the bound state via tuning the Larmor frequency.

\section{Acknowledgments}We are grateful to Hao-Wen Su for his
valuable discussions. This work is supported by the National Natural Science Foundation of China (Grants No. 62461160263, No. 12275109, and No. 92576202), Quantum Science and Technology-National Science and Technology Major Project (Grants No. 2023ZD0300200 and No. 2023ZD0300904), the Natural Science Foundation of Gansu Province (Grants No. 26RCKA011 and No. 25JRRA799), and Guangdong Provincial Quantum Science Strategic Initiative (Grant No. GDZX2505004).

\bibliography{ref}
\end{document}